\documentstyle[aps,prl,psfig]{revtex}

\begin{document}
\draft

\twocolumn[\hsize\textwidth\columnwidth\hsize\csname@twocolumnfalse\endcsname

\author{Dario Alf\`e}

\address{ Department of Earth Sciences, and Department of Physics and
Astronomy, \\ University College London, Gower Street, London, WC1E
6BT, U.K }

\title{ First principles simulations of direct coexistence of solid
and liquid aluminium }

\maketitle

\begin{abstract}
First principles calculations based on density functional theory, with
generalised gradient corrections and ultrasoft pseudopotentials, have
been used to simulate solid and liquid aluminium in direct coexistence
at zero pressure. Simulations have been carried out on systems
containing up to 1000 atoms for 15 ps. The points on the melting curve
extracted from these simulations are in very good agreement with
previous calculations, which employed the same electronic structure
method but used an approach based on the explicit calculation of free
energies~\cite{vocadlo02}.
\end{abstract}

\pacs{PACS numbers: 
64.10.+h 
64.70.Dv 
71.15.-m 
71.15.Pd  
}
]

The calculation of melting properties of materials using computer
simulations has a long history. Two main approaches have traditionally
been used. The first is based on the direct simulation of solid and
liquid in
coexistence~\cite{morris94,laio00,belonoshko00,belonoshko00b,jesson2000a}.
The second on the calculation of the free energy of solid and
liquid~\cite{moriarty84,broughton87,mei92,straub94,jesson2000}, with
the melting point $(p,T)$ determined by the condition of equality of
the Gibbs free energies of liquid and solid, $G_{\rm liq}(p,T) =
G_{\rm sol}(p,T)$. The two approaches must clearly give the same
answer once all the sources of errors have been brought under
control. These works were originally carried out using classical model
potentials, with parameters usually adjusted to reproduce some known
experimental properties of the material, or fitted to {\em ab-initio}
calculations. The advantage of using these classical potentials is
that they are relatively simple, so that computer simulations could
easily be carried out on large systems and for long time. The main
disadvantage, however, is that the transferability of these potentials
is not always guaranteed, so that the accuracy of the predictions is
sometimes questionable.  In 1995 Sugino and Car~\cite{sugino95} showed
that it was actually possible to use density functional theory
(DFT)~\cite{car85} techniques to calculate the melting temperature of
materials truly from first principles, i.e without relying on any
adjustable parameters. They chose the melting of silicon as a test
case, and using the local density approximation (LDA) they calculated
the Gibbs free energy of solid and liquid using thermodynamic
integration. This is a well known statistical mechanics technique to
compute free energy differences between two
systems~\cite{frenkel96}. The idea is to use a simple system for which
the free energy is known as a reference system, and then compute the
free energy difference between the {\em ab-initio} and the reference
system, which is equal to the reversible work done in adiabatically
switching the potential energy from one system to the other. An
attractive feature of this method is that the final result is totally
independent on the choice of the reference system.  In practice,
however, this method can only work if one is able to find a reference
system which is as close as possible to the {\em ab initio } system,
so that the computational effort needed to calculate the free energy
difference between the two systems is reduced to the minimum.  Using
this method,
Sugino and Car found a zero pressure melting temperature about 20\%
lower than the experimental datum. This may seem not particularly
good, but it has to be remembered that no adjustable parameters were
inputed in the calculations, nor any experimental data other than the
Plank's constant and the mass and charge of the electron.  Recently,
we have argued~\cite{alfe03} that this non-perfect agreement between
LDA results and experiments is probably due to non-cancelling LDA
errors between the solid and liquid. The reason is that the two phases
have very different properties: the liquid is six-fold coordinated and
the solid four-fold coordinated, the liquid is a metal and the solid
an insulator, so it is not surprising that the LDA errors in the two
phases may be different. We have repeated the LDA calculations of
Sugino and Car using similar techniques and found the same melting
temperature. However, we have also shown that if the generalised
gradient corrections (GGA) are used, then the melting temperature of
silicon comes out in much closer agreement with the
experiments~\cite{alfe03}.

Since the work of Sugino and Car~\cite{sugino95} a number of first
principles based calculations of melting points and melting curves
have followed. In some cases the free energy approach has been
used~\cite{dewijs98,alfe99,alfe01,alfe02,vocadlo02}, while other
methods relied on fitting a model potential to first principles
simulations and calculating melting properties with the model
potential~\cite{laio00,belonoshko00,belonoshko00b}.  The advantage of
the free energy approach is that it is unbiased, provided all sources
of technical errors are brought under control. A disadvantage of the
method is that it is intrinsically complex. On the other hand, the
coexistence approach is relatively simple to apply, but has the main
disadvantage of relying on good quality fitting and, more importantly,
transferability of the model for at least a simultaneous good
description of solid and liquid. Recently, we have shown that provided
that the model potential is reasonably close to the {\em ab-initio}
system~\cite{small_fluctuations}, it is possible to correct for the
(small) differences between the model and the full {\em ab-initio}
system using a perturbational approach to thermodynamic integration,
and that once the corrections are applied the results coincide with
those obtained using the free energy approach~\cite{alfe02b}.

The coexistence approach has been used extensively in the past to
calculate the melting properties of various materials. In the constant
volume constant internal energy ($NVE$ ensemble) approach to the
method it has been shown that liquid and solid can coexist for long
time, provided $V$ and $E$ are appropriately
chosen~\cite{morris94,alfe02b}. The average value of the pressure $p$
and temperature $T$ over the coexisting period then give a point on
the melting curve. Size effects have also been studied quite
extensively, and it was shown that correct results can be obtained in
systems containing more than 500
atoms~\cite{belonoshko00b,morris02}.

Here I have exploited recent advances in computer power and algorithms
developments to use direct DFT calculations for simulating solid and
liquid aluminium in coexistence near zero pressure. This work combines
the simplicity of the coexistence approach with the accuracy provided
by DFT, and in some respect represents a shift of paradigm, whereby
the main effort is transferred from the human to the computer. The
temperature at which solid and liquid coexist are found to be in good
agreement with our previous results obtained with the free energy
approach~\cite{vocadlo02}, so that these results also provide
additional evidence that our techniques to calculate free energies are
sound.

The calculations have been performed with the VASP
code~\cite{kresse96}, with the implementation of an efficient
extrapolation of the charge density~\cite{alfe99e}, ultrasoft
pseudopotentials~\cite{vanderbilt90} with a plane wave cutoff of 130
eV, and generalised gradient corrections.  The simulations have been
performed in the $NVE$ ensemble on systems containing 1000 atoms
($5\times 5\times 10$ cubic supercell), with a time step of 3 fs and a
convergency threshold on the total energy of $2\times
10^{-7}$~eV/atom. With these prescriptions the drift in the
microcanonical total energy was less than $0.3$~K/ps.  The total
length of the simulations were typically 15 ps.
The volume per atom was chosen to be $V=18.5$~\AA$^3$, which is close
to the average of the volumes of solid and liquid at the zero pressure
melting point calculated in our previous work~\cite{vocadlo02}.
Electronic excitations have been included within the framework of
finite temperature DFT.  The simulations have been performed using the
$\Gamma$ point only, and spot checked with
Monkhorst-Pack~\cite{monkhorst76} ($2\times 2\times 1$) and ($4\times
4\times 2$) ${\bf k}$-point grids.
Calculations with the $\Gamma$ point predict a small non-hydrostatic
stress tensor with a difference of about $p_z - (p_x+p_y)/2 = -2$~kB
between the components of the stress parallel to the solid-liquid
interface, $p_x$ and $p_y$, and that perpendicular to the interface,
$p_z$. The pressure $p = (p_x + p_y + p_z)/3$ is essentially exact
and the three off-diagonal components of the stress tensor fluctuate
around zero average, so that there is no shear stress on the cell.
With $\Gamma$ the total energy is wrong by $\approx 5$~meV/atom, and
it is likely that the error is almost equally shared by the liquid and
the solid parts, so that the resulting error on the melting
temperature is most probably negligible~\cite{delta_t}. A correction
term of 2.7 kB due to the lack of convergency with respect to the
plane wave cutoff has been added to the calculated pressures.  The
systems have been monitored by inspecting the average number density
in slices of the cell taken parallel to the boundary between solid and
liquid. In the solid region this number is a periodic function of the
slice number, and in the liquid part it fluctuates randomly around its
average value.


The zero pressure crystal structure of aluminium is face-centred-cube
(fcc), so I have assumed that melting occurs from this structure.  To
prepare the system I have used the inverse power classical potential
employed in Ref.~\cite{vocadlo02}. This model has been tuned to the
same {\em ab-initio} system used here, so it is a good starting point
for the present calculations. The preparation procedure follows
Ref.~\cite{alfe02b}. A perfect crystal is initially thermalized at 800
K, then the simulation is stopped, half of the atoms are clamped and
the other half are freely evolved at very high temperature until
melting occurs, then the liquid is thermalized back at 800
K~\cite{thermalize}. At this point the system is being freely evolved
in the $NVE$ ensemble using DFT.

\begin{figure}
\psfig{figure=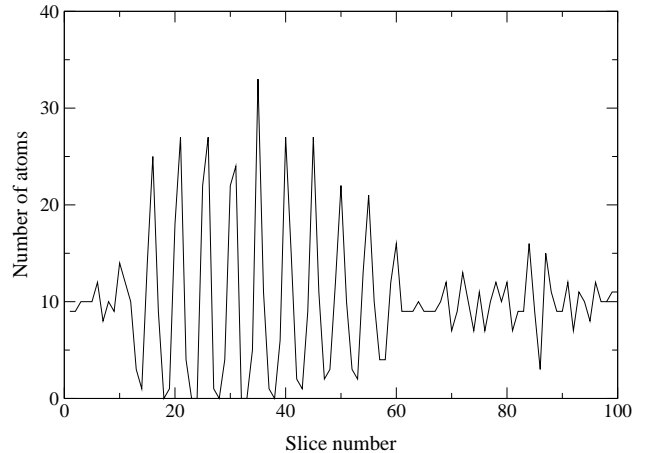,height=2.6in,angle=-90}
\caption{Density profile in a simulation of solid and liquid Al
coexisting at zero pressure. The system is divided in slices of equal
thickness (0.42 \AA) parallel to the solid-liquid interface, and the
graph shows number of atoms in each slice. Simulations were performed
on a system of 1000 atoms using density functional theory.}~\label{fig:dens}
\end{figure}
\begin{figure}
\psfig{figure=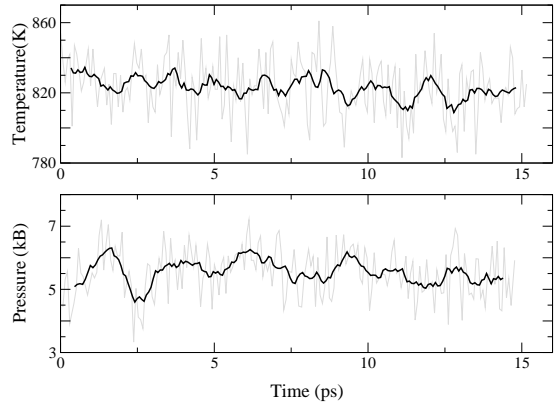,height=2.6in,angle=-90}
\caption{Time variation of temperature (upper panel) and pressure
(lower panel) during a simulation of solid and liquid Al in
coexistence. Simulations were performed on a system of 1000 atoms with
density functional theory. Gray lines: actual data; black lines:
running averages over a 0.75 ps period.}~\label{fig:pt}
\end{figure}
\begin{figure}
\psfig{figure=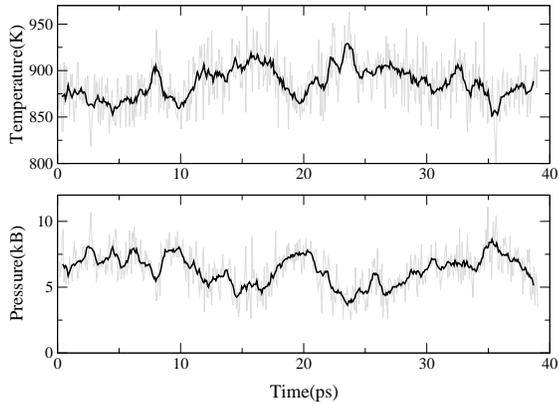,height=2.6in,angle=-90}
\caption{Time variation of temperature (upper panel) and pressure
(lower panel) during a simulation of solid and liquid Al in
coexistence for a system containing 512 atoms. Gray lines: actual data;
black lines: running averages over a 0.75 ps period.}~\label{fig:pt512}
\end{figure}

As explained in Ref.~\cite{alfe02b}, for each chosen volume there is a
whole range of internal energies for which different amount of solid
and liquid are in coexistence. Provided that the energy is not too low
(high) so that that the whole system freezes (melts), a whole piece of
melting curve can be obtained for any fixed volume. In this work I
have performed 3 simulations with 1000 atoms, all at the same volume
but with different amounts of total energy, which therefore provide
three distinct points on the melting curve.
In all simulations coexistence was obtained for the whole length of
the runs.  In Fig.~\ref{fig:dens} I display the density profile,
calculated by dividing the simulations cell into 100 slices parallel
to the solid-liquid interface, corresponding to the last configuration
of one of these simulations.  Fig.~\ref{fig:pt} contains the calculated
temperature (upper panel) and pressure (lower panel) for this
particular simulation, and shows that they oscillate stably around
their average values $T\approx 820$~K and $p\approx
5.5$~kB~\cite{continuation}.  In order to test the reliability of the
lengths of these simulations I have performed additional runs on
system containing 512 atoms ($4\times 4\times 10$ cubic supercell),
which could be simulated for up to 40 ps. Temperature and pressure
from one of these runs are displayed in Fig.~\ref{fig:pt512}. It is
clear that all the information needed to extract useful values for $p$
and $T$ is contained in any time window of $\approx 5-10$ ps, which
provides confidence that the simulations for the large systems are
long enough.  A more detailed inspection of the figures reveals the
presence of anti-correlated oscillations in pressure and temperature,
which correspond to fluctuations in the total amount of solid and
liquid in the system. These fluctuations seem absent in the
simulations with 1000 atoms, but they would probably develop if the
runs could be extended. In any case, they do not affect the average
value of $p$ and $T$.

It has been pointed out that non-hydrostaic conditions artificially
raise the free energy of the solid, and therefore lower the melting
temperature~\cite{morris02}.  To investigate the effect on the melting
temperature due to the present non-hydrostatic conditions I have used
the model potential used in Ref.~\cite{vocadlo02}. The model showed a
difference $p_z - (p_x+p_y)/2 = 0.5$~kB when simulated with the same
cell as the {\em ab-initio} system with 1000 atoms. To estimate the
systematic error on the melting temperature I have performed an
additional simulation using a slightly elongated cell at the same
volume, so that non-hydrostaticity was reduced to less than 0.1
kB. The decrease of melting temperature in this second simulation was
less than 10 K. A similar effect is expected for the {\em ab-initio}
system, so that I estimate a systematic error due to non-hydrostatic
conditions to be at worse of the order of 20 K.

\begin{figure}
\psfig{figure=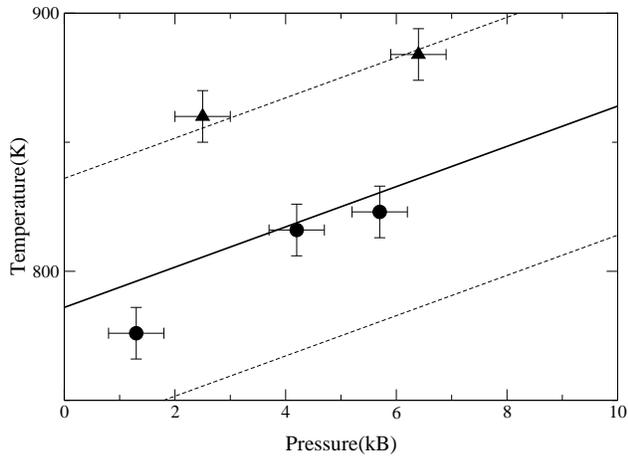,height=2.6in,angle=-90}
\caption{Temperatures and pressures at which liquid and solid coexist
in simulations containing 1000 atoms (circles) and 512 atoms
(triangles). The solid line is the lower end of the melting curve
calculated using the free energy approach in
Ref.~\protect\cite{vocadlo02} (see text), light dashed lines represent
error bars.  }~\label{fig:melt}
\end{figure}

In Fig.~\ref{fig:melt} I report the values of $p$ and $T$ calculated
here in comparison with the low end of the melting curve calculated in
Ref.~\cite{vocadlo02} The agreement between the present finding with
1000 atoms and the free energy approach results is extremely good. The
results obtained from the simulations with 512 atoms display melting
temperatures higher by about 50 K, although they are still compatible
with the results obtained with the free energy approach within the
combined error bars. In order to test if the cause of this size effect
was due to inadequate {\bf k}-point sampling, I have performed a
simulation using a $(2\times 2\times 1)$ {\bf k}-point grid on a
system with 512 atoms for $\approx 6$ ps. The length of this
simulation is too short to draw definite conclusions, but it indicates
that the pressure may be underestimated by $\approx 1-2$~kB and the
temperature over-predicted by $\approx 20$~K. This would bring the
results in somewhat better agreement with those obtained with the
larger system, and also with the calculations based on the free energy
approach (which were fully converged with respect to size and {\bf
k}-point sampling)~\cite{vocadlo02}.  Notice that these results do not
agree perfectly with the experimental zero pressure melting
temperature of aluminium (933 K)~\cite{crc97}. We have argued in our
previous work~\cite{vocadlo02} that this disagreement is due to
inadequacy of the GGA to predict the vibrational properties of the
solid. We have also suggested that this inaccuracy can be rationalised
in terms of the errors in the equation of state, and devised a simple
way to correct for it. In Table 1 I report the calculated GGA lattice
constant, the bulk modulus and the cohesive energy compared with the
experimental data.  Following Gaudoin et al.~\cite{gaudoin02}, I also
report the experimental values adjusted for the absence of zero point
motion effects, which is missing in the present calculated values. It
is evident that the GGA predicts a zero pressure lattice constant for
Al which is too large, or, which means that at the correct lattice
constant GGA predicts a positive pressure of about 16 kB. It follows
that the GGA zero pressure melting point is actually the melting point
at a negative pressure of about -16 kB, which is about 120-130 K
lower.


In summary, I have shown here that it has become possible to directly
simulate solid and liquid in coexistence using density functional
theory techniques. This method combines the advantages of being
relative easy to use and provide DFT accuracy. It is inevitably
computationally very intensive, and these calculations have only been
possible thanks to an access onto the 3.5 Teraflop/s machine (IBM p690
Regatta with 1280 processors). The cost of each molecular dynamics
step for the 1000 atoms system was between 3 and 4 minutes on 128
processors of the machine. However, as large computer resources become
routinely available this method should find widespread applicability
in the future. The present results are in very good agreement with our
previous findings based on the direct calculations of free energies,
and therefore also support the reliability of those techniques.

This work has been supported by the Royal Society and by the
Leverhulme Trust. The author wishes to acknowledge the computational
facilities of the UCL HiPerSPACE Centre (JREI grant JR98UCGI), and the
early access to the HPC{\em x} computer facilities in Daresbury,
granted by NERC through support of the Mineral Physics Consortium. The
author is grateful to C. Wright for computational technical support
and to M. J. Gillan and G. D. Price for discussions. The author also
wishes to thank M. Pozzo for useful suggestions on the manuscript.

\begin{table}
\begin{tabular}{l|ll}
  &  Experiment  &  GGA  \\
\hline 
$E$ & 3.39 (3.43) & 3.43 \\
$B$ & 76 (81) & 73  \\
$a_0$  & 4.049 (4.022) & 4.05 \\
\end{tabular}
\caption{ Experimental values for the cohesive energy
$E^c$~(eV/atom)~\protect\cite{kittel}, the bulk modulus
$B$~(GPa)~\protect\cite{crc97} and the lattice constant
$a_0$~(\AA)~\protect\cite{crc97}. In parenthesis are the adjusted
values for the effect of zero point motion~\protect\cite{gaudoin02}.
The calculated GGA values have been obtained from a fit to a
Birch-Murnaghan equation of state~\protect\cite{birch47}, and they do
not include zero point motion effects.}

\end{table}

\begin{thebibliography}{99}
\bibitem{delta_t} Even in the unlikely worse case scenario in which
this error was borne entirely by the solid (or entirely by the
liquid), this would result in a wrong prediction of the melting
temperature by only about 30~K.
\bibitem{thermalize} The temperature of 800~K has been chosen because
it is close to the DFT-GGA melting temperature previously
calculated~\cite{vocadlo02}.
\end{thebibliography}
\end{document}